\documentclass[aps,prl,amssymb,showpacs,preprint]{revtex4}
\usepackage{graphicx}
\usepackage{bm}

\begin{document}

\title{Polyamorphism of ice at low temperatures from constant-pressure
  simulations}
\author{R.~Marto\v{n}\'ak} 
\altaffiliation[Permanent address: ]{Department of Physics (FEI), 
  Slovak University of Technology, Ilkovi\v{c}ova 3, 812 19 Bratislava,
  Slovakia} 
\email{martonak@phys.chem.ethz.ch}
\affiliation{Computational Science, Department of Chemistry and Applied
  Biosciences, ETH Zurich, USI Campus, Via Giuseppe Buffi 13, CH-6900
  Lugano, Switzerland} 
\author{D.~Donadio}
\affiliation{Computational Science, Department of Chemistry and Applied
  Biosciences, ETH Zurich, USI Campus, Via Giuseppe Buffi 13, CH-6900
  Lugano, Switzerland} 
\author{M.~Parrinello}
\affiliation{Computational Science, Department of Chemistry and Applied
  Biosciences, ETH Zurich, USI Campus, Via Giuseppe Buffi 13, CH-6900
  Lugano, Switzerland} 
\date{\today}

\begin{abstract}
  We report results of MD simulations of amorphous ice in the pressure
  range 0 -- 22.5 kbar. The high-density amorphous ice (HDA) prepared by
  compression of I$_{\rm h}$ ice at $T=80$ K is annealed to $T=170$ K at
  intermediate pressures in order to generate relaxed states. We confirm
  the existence of recently observed phenomena, the very high-density
  amorphous ice and a continuum of HDA forms. We suggest that both
  phenomena have their origin in the evolution of the network topology of
  the annealed HDA phase with decreasing volume, resulting at low
  temperatures in the metastability of a range of densities.
\end{abstract}

\pacs{64.70.Kb, 61.43.Er, 02.70.Ns, 07.05.Tp} 
\maketitle 

Ice at low temperatures exhibits complex behavior. Besides many crystalline
forms two amorphous ones have been known for a long time.  High-density
amorphous ice (HDA) is prepared by compression of ordinary I$_{\rm h}$ ice
to 12 kbar \cite{mishima1} and when recovered at ambient pressure it has a
density of $\sim$ 1.17 g/cm$^3$.  Upon isobaric heating to 117 K, the
density drops considerably and a second distinct form is found, called
low-density amorphous ice\cite{mishima1} (LDA) with a density $\sim$ 0.95
g/cm$^3$. The transition between LDA and HDA can also be induced by
pressure \cite{mishima2,mishima3}. The existence of two different amorphous
forms of ice is crucial for the hypothesis of two kinds of supercooled
water\cite{twoliquid} where LDA and HDA represent a continuation of the two
distinct liquid phases below the freezing point\cite{pooletal}; in this
context the apparent sharpness of the transition between HDA and
LDA\cite{mishima2,mishima3} is of special importance. Recently several new
experimental results\cite{loerting,klotz,finney_vhda,tulk,guthrie,mishima4}
raised new questions about the polyamorphism of ice.

A new amorphous form of ice has been reported \cite{loerting}, prepared by
heating HDA under pressure of 11 kbar to $T \sim 170$ K and cooling it back
to $T=80$ K; when recovered at ambient pressure it has a density of $\sim$
1.25 g/cm$^3$. It has been called very high-density amorphous ice (VHDA)
and characterized experimentally using neutron diffraction
\cite{finney_vhda}. Moreover, experiments\cite{tulk, guthrie} show that by
heating HDA to temperatures intermediate between 80 K and 110 K the sample
can be trapped in an apparent continuum of metastable structures between
HDA and LDA.  This suggests that there might be no sharp transition between
the two forms.  On the other hand, Ref.\cite{mishima4}, while also finding
a continuum of HDA states, observed a propagation of the LDA-HDA interface,
thus still favoring a sharp transition between the two forms.  Possible
implications of the new experiments have been discussed\cite{klug,soper}.

Here we suggest that both new phenomena, the VHDA and the continuum of HDA
densities, originate from a relation between the density and the topology
of the hydrogen-bonded network of the HDA phase.  Our tool is
constant-pressure molecular dynamics (MD) simulation\cite{ft8}, using the
classical TIP4P model\cite{tip4p}. This was found to reproduce well the
transitions I$_{\rm h}$ -- HDA and LDA -- HDA, both qualitatively and
quantitatively\cite{klein,pooletal,okabe}.  Our system contained 360 H$_2$O
molecules and the total accumulated simulation time is more than 5
$\mu$s\cite{ft1}.

\begin{figure}[h]
\includegraphics*[width=12cm]{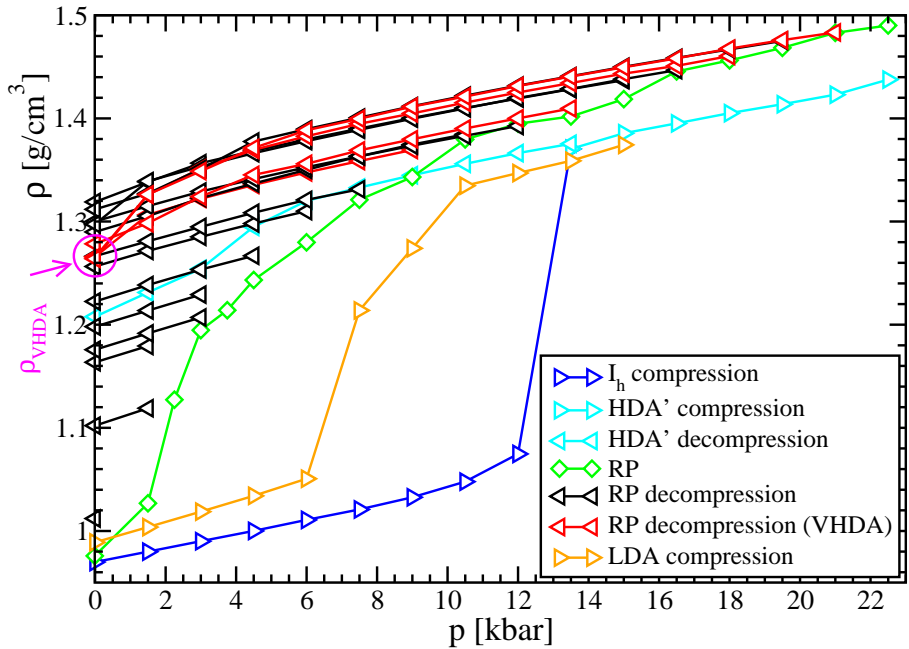}
\caption{Density vs.~pressure dependence at $T=80$ K for the various
  amorphous phases during compression/decompression.  The triangles point
  in the direction of pressure change, the lines are just a guide for the
  eye.}
\label{fig_ro_p}
\end{figure}

\begin{figure}[h]
  \includegraphics*[width=12cm]{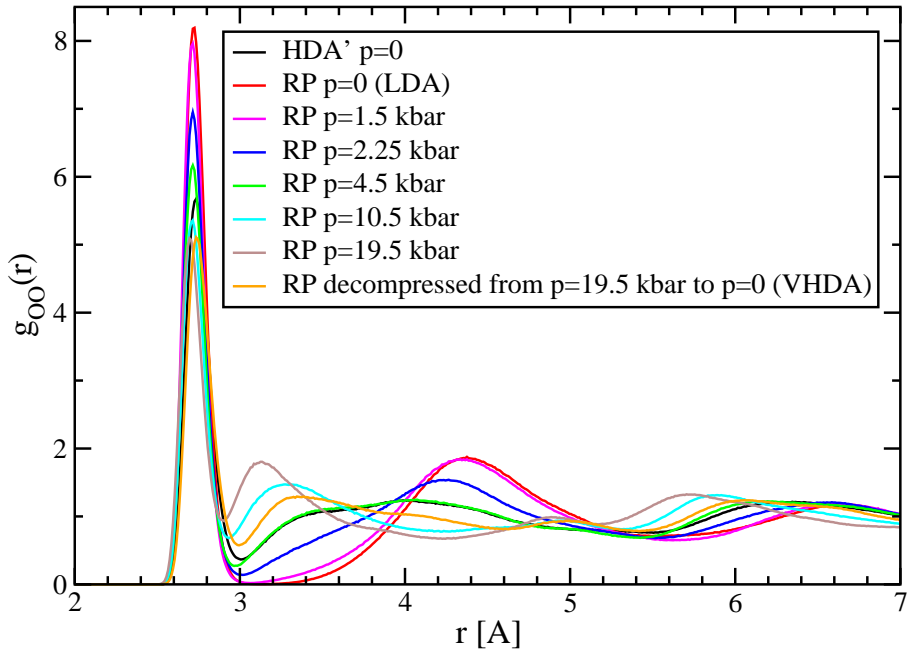}
\caption{Oxygen -- oxygen radial distribution function of various amorphous
  phases at $T=80$ K and $p = 0 - 19.5$ kbar. }
\label{fig_rdf}
\end{figure}

We first compressed the I$_{\rm h}$ ice at $T = 80$ K, increasing the
pressure in steps of 1.5 kbar.  At 13.5 kbar a sharp transition occurs and
the density increases by almost 30 \% to 1.37 g/cm$^3$
(Fig.\ref{fig_ro_p}). The sample was then further compressed at $T=80$ K to
22.5 kbar and from 15 kbar decompressed to $p=0$; we call this as-prepared
HDA phase HDA'.  During decompression the HDA' density gradually decreased
and at $p = 0$ reached the value of 1.21 g/cm$^3$, close to the HDA
experimental value of 1.17 g/cm$^3$\cite{mishima1}.  The radial
distribution function (RDF) of HDA' at $p=0$ is shown in Fig.\ref{fig_rdf};
it has a broad second peak between 3.3 and 4.6 \AA, very similar to that of
HDA at $p=0$\cite{klotz}.  Inspired by the experiments\cite{loerting} that
led to the discovery of the VHDA we decided to anneal the HDA' phase at
each intermediate pressure between 22.5 kbar and zero\cite{ft4} in order to
search for possible new structures.  Annealing was performed by heating up
to $T=170$ K and subsequent cooling down to 80 K; the temperature was
always changed in steps of 10 K.  The phase obtained in this way will be
called relaxed phase (RP).  While in experiment an HDA' sample heated at an
arbitrary pressure might recrystallize\cite{pooletal,klug,klotz_recryst},
in the time scale of a simulation this is not likely to happen. We are
therefore restricted to exploring the (metastable) disordered structures.
The RP phase prepared at each pressure was afterwards decompressed at
$T=80$ K down to $p=0$, decreasing the pressure in steps of $1.5$ kbar.

\begin{figure}[h]
  \includegraphics*[width=12cm]{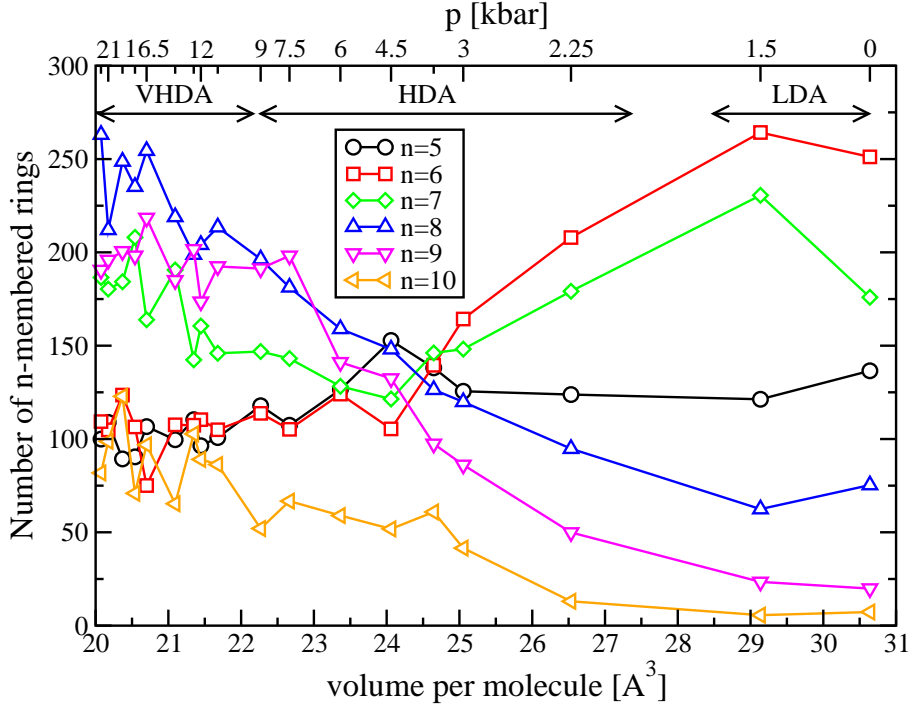}
\caption{Number of $n$-membered network rings in the RP as function of
  volume at $T=80$ K and $p = 0 - 22.5$ kbar. }
\label{fig_rings}
\end{figure}

The O-O RDF's of RP at different pressures are shown in Fig.\ref{fig_rdf}.
We also calculated the O-H RDF (not shown) for RP at $p=0, 2.25, 6$ and
$15$ kbar. Integrating between 1.5 and 2.25 \AA~we found at all pressures a
coordination number of 2, indicating a fully hydrogen-bonded network.  In
order to characterize the evolution of the network topology we calculated
the ring statistics \cite{rings} for RP at all pressures. This is able to
reveal information on medium-range order, which might not be easily
extracted from the RDF\cite{trachenko,davila}. We now discuss the rather
remarkable behavior of the RP at increasing pressure in terms of density,
RDF and network ring statistics (Fig.\ref{fig_rings}) and show that there
are 3 distinct regimes.

At $p = 0$ the density after annealing reaches a value of 0.98 g/cm$^3$.
This agrees well with the experimental LDA value of 0.95 g/cm$^3$. The RDF
of this phase (Fig.\ref{fig_rdf}) exhibits at $r = 3.1$ \AA~a very deep
minimum between the first and second shell and a well-defined second shell
peak at $r = 4.4$ \AA, very similar to that found experimentally for LDA in
Ref.\cite{finney_hlda}. The RP at $p = 0$ thus coincides with the LDA as
expected.  At 1.5 kbar the density increases to 1.03 g/cm$^3$ while the RDF
remains very similar to that of LDA at $p=0$, with the very deep minimum
between the first and second shell; at both pressures the network is
dominated by 6-membered rings.  A substantial number of 7-membered rings is
also present and upon change of pressure from 0 to 1.5 kbar this increases
and approaches the number of 6-membered rings, while a smaller number of 5,
8 and 9-membered rings stays almost constant.  The LDA phase from 1.5 kbar
relaxes at $p=0$ to $\rho = 1.01$ g/cm$^3$, close to that of LDA prepared
at $p=0$.

The properties of the RP at $p=2.25$ kbar are rather different. The density
increases much faster and reaches a value of 1.13 g/cm$^3$. The second
shell peak of RDF drops and shifts to lower $r$ and at the same time RDF
grows substantially in the region around $r = 3.3$ \AA, revealing the
presence of interstitial molecules\cite{klotz}. A dramatic change is seen
in the ring statistics: the number of 6 and 7-membered rings now drops fast
and at the same time the number of 8 and 9-membered rings grows fast.  This
behavior is compatible with a transition from LDA to HDA occurring between
1.5 and 2.25 kbar; our resolution does not allow us to determine whether
there is a discontinuous jump in the density of RP between 1.5 and 2.25
kbar. Around $p=4.5$ kbar the density growth slows down and RDF develops a
broad second peak between 3.2 and 4.4 \AA, quite similar to that of HDA' at
$p=0$ (Fig.\ref{fig_rdf}).  Approaching $p \sim 9$ kbar the ring statistics
are definitely dominated by 8 and 9-membered rings; the network has thus
undergone a substantial reconstruction\cite{silica}. When decompressed to
$p=0$, the structures from $p=2.25 - 9$ kbar relax to densities between
1.10 and 1.26 g/cm$^3$; the decompression curves (black curves in
Fig.\ref{fig_ro_p}) are roughly parallel (also to the I$_{\rm h}$ and LDA
compression curves below the respective transition pressures) and change
slowly with pressure, indicating in all cases a similar compressibility,
much lower than that of the RP.

For $p > 9$ kbar, the number of 6 and 9-membered rings stabilizes,
indicating that the network reconstruction has been partially completed;
only the number of 7 and 8-membered rings continues to grow\cite{networks}.
At the same time the density growth slows down further and the RDF develops
a pronounced second peak at $r = 3.2$ \AA~(Fig.\ref{fig_rdf}) while the
original second shell peak at $r = 4.4$ \AA~disappears completely.  Upon
decompression from pressures between 9 and 22.5 kbar the density of the RP
converges to somewhat higher values ranging from 1.26 g/cm$^3$ to 1.31
g/cm$^3$. Several of these decompression curves, including the one from the
highest pressure 22.5 kbar, however, upon decreasing pressure bend down
more than others (red curves in Fig.\ref{fig_ro_p}) and reach at $p=0$
practically the same density of about 1.26 -- 1.28 g/cm$^3$.  It is then
plausible to assume that the curves which stay at slightly higher density
do so just because of our short observation time and upon a longer
simulation they would all reach the same point.  We note that in
Ref.\cite{loerting} the samples annealed at 11 and 19 kbar reached the same
density upon decompression.  The clear similarity of the RDF of RP
decompressed from 19.5 kbar to that of VHDA recovered at $p=0$ in
experiment\cite{finney_vhda} manifested in the distinct peak at $r=3.37$
\AA, very close to the first shell peak, allows us to identify this form as
VHDA.  Our results are thus compatible with the existence at $p=0$ of a
well-defined LDA state and a continuum of metastable HDA states with
density below that of VHDA\cite{mishima4}.

\begin{figure}[h]
  \includegraphics*[width=12cm]{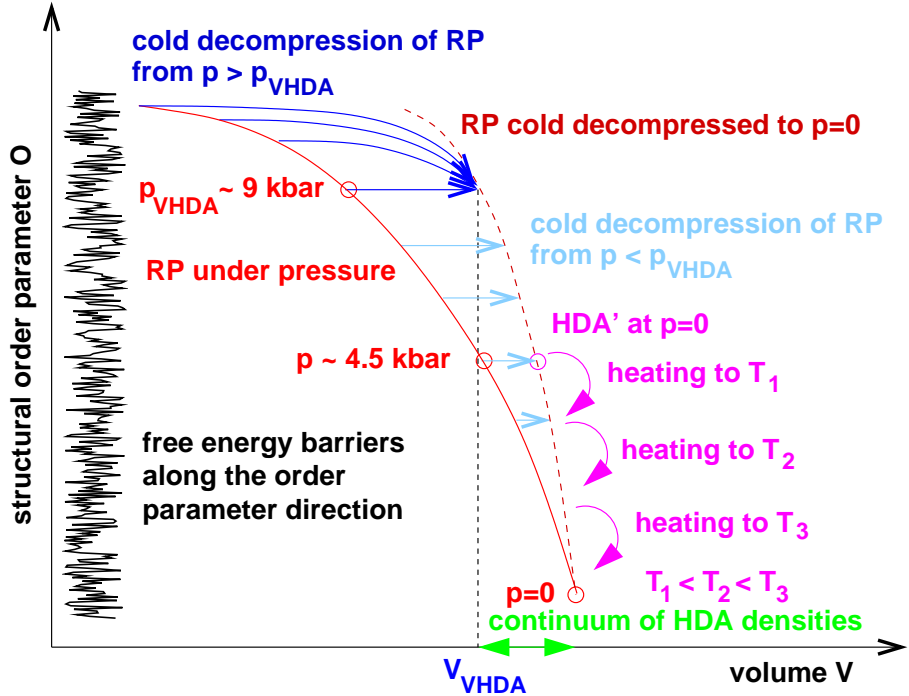}
\caption{Schematic picture of decompression of RP from different pressures
  in the $V$ -- $O$ plane. The structural order parameter $O$ is for
  simplicity shown as one-dimensional.}
\label{fig_ord_vs_v}
\end{figure}

In order to interpret the observed behavior we represent all HDA structures
in the $V$ -- $O$ plane (Fig.\ref{fig_ord_vs_v}), where $V$ is the volume
and $O$ is a structural order parameter related to network topology (e.g.,
the number of 6 and 8-membered rings). As discussed above
(Fig.\ref{fig_rings}), $O$ changes more slowly at small values of $V$ and
faster at large values of $V$; this is represented by the downward
curvature of the RP curve in Fig.\ref{fig_ord_vs_v}. The location of the
$p=0$ point of RP represents the global minimum of the free energy
$F_{HDA}(V)$ of the HDA branch.  RP is able to change its density with
pressure in a substantial way due to the accompanying network
reconstruction. This must involve breaking and remaking of bonds, which at
any volume requires crossing a free energy barrier. Along the vertical
direction of $O$ the free energy surface is therefore very rough, with a
manifold of local minima. We now analyze how RP prepared at a certain
pressure relaxes when decompressed at cold conditions ($T=80$ K) to $p=0$.
An elastic relaxation of the compressed network, without reconstruction,
corresponds to horizontal transition at constant $O$. All such metastable
minima fall at the dashed red curve in Fig.\ref{fig_ord_vs_v}. At small $p$
and large $V$, a further increase of volume would require a large change of
$O$, resulting in a higher barrier.  At the same time the gain of free
energy is small since approaching its minimum the $F_{HDA}(V)$ curve
becomes increasingly flat; vertical relaxation of $O$ is therefore
impossible and the system at cold conditions remains trapped in a continuum
of metastable minima.  This is the case for the black curves in
Fig.\ref{fig_ro_p}. We note that HDA' as prepared under strongly
non-equilibrium conditions by low-temperature pressure-induced
amorphization of I$_{\rm h}$ ice is likely to be located away from the RP
curve.  Upon decompression to $p=0$, however, it should also approach some
point on the dashed red curve.  In fact, the density as well as the RDF of
our RP decompressed from 4.5 kbar (not shown) is very similar to that of
HDA' at $p=0$.  A relaxation of $O$ then becomes possible upon progressive
increase of the temperature\cite{tulk,guthrie,mishima4}; the system
approaches the global minimum of $F_{HDA}(V)$ and at a certain point
undergoes a transition to a more stable LDA form.  As observed by X-ray
scattering in Ref.\cite{tulk,guthrie}, with increasing annealing
temperature the interstitial region around 3.5 \AA~is depleted and the
second shell peak around 4.4 \AA~grows. The same qualitative trend is
observed in our RDF's of RP decompressed from $p = 3 - 4.5$ kbar (not
shown) upon decreasing $p$.  It is thus plausible to assume that by
annealing HDA' under pressure and subsequent cold decompression of the RP
in simulation we have reached the region of states which experimentally can
be accessed by progressive annealing of HDA' at $p=0$.

The above mechanism for stabilizing decompressed metastable states ceases
to be operative at higher $p$ and low $V$. Here an increase of volume is
associated with just a small change of $O$ resulting in a smaller barrier;
the possible gain in free energy is larger because the $F_{HDA}(V)$ curve
is steep. The probability of crossing the barrier is thus enhanced and a
change of $O$ can occur even at $T=80$ K.  This suggests the existence of a
pressure $p_{VHDA}$ such that for $p > p_{VHDA}$ the RP upon decompression
can relax with a change of $O$ and reach a certain lowest metastable volume
$V_{VHDA}$.  This corresponds to the behavior of the red curves in
Fig.\ref{fig_ro_p}, indicating $p_{VHDA} \sim 9$ kbar and $V_{VHDA}$
equivalent to density $\rho_{VHDA} \sim$ 1.27 g/cm$^3$.  The origin of the
VHDA might thus be kinetic rather than thermodynamic.

\begin{figure}[h]
  \includegraphics*[width=12cm]{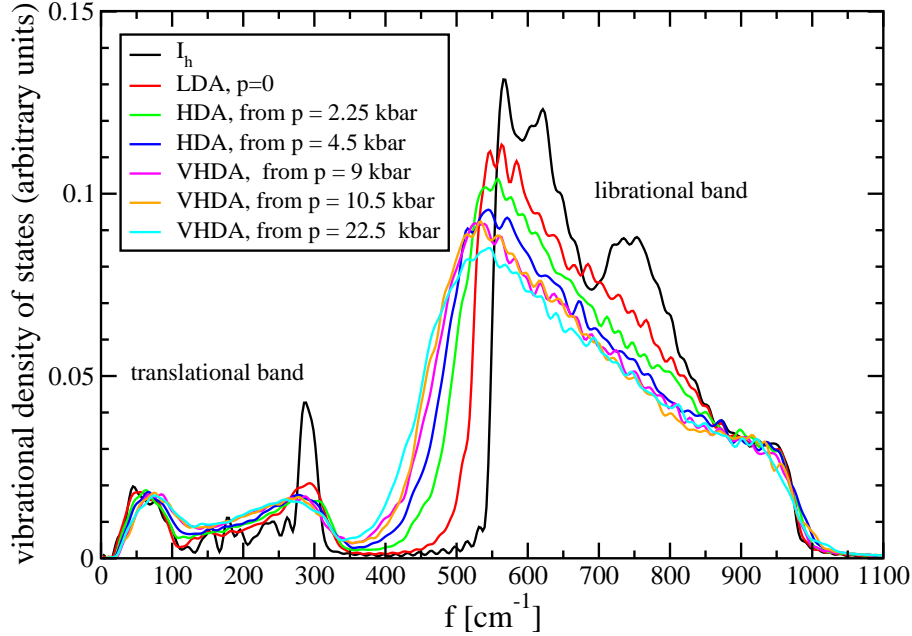}
\caption{VDOS for hydrogen atoms (Fourier transform of the
  velocity-velocity autocorrelation function) in the decompressed RP.
  I$_{\rm h}$ ice is also shown for comparison.}
\label{fig_dens_vib}
\end{figure}

We calculated the vibrational density of states (VDOS) for hydrogen atoms
(Fig.~\ref{fig_dens_vib}) in the RP decompressed from various pressures.
The main trend is the pronounced shift of the band edge of the librational
band to lower frequencies upon increasing density. This could be checked
experimentally on the annealed samples with continuously varying
density\cite{tulk}.  It might also be possible to identify in vibration
spectra direct fingerprints of network rings with specific size, as done
for amorphous silica\cite{pasquarello}.

In conclusion, in this paper we reproduced the new experimental facts, the
VHDA and the continuum of HDA states at $p=0$.  Both phenomena appear to
originate from the continuous evolution of the network topology of the HDA
branch of the RP phase under pressure, resulting in a metastability of a
range of HDA densities at $p=0$ and $T=80$ K. As revealed by the network
ring statistics, HDA between 2.25 and 9 kbar represents a gradual evolution
of LDA into VHDA; however, the derivatives of the number of rings with
respect to volume change abruptly between LDA and HDA. This indicates that
while the density difference between LDA and HDA might be smaller than
previously thought there is a clear topological difference between the two
phases.  Therefore the existence of HDA and LDA is still compatible with
the hypothesis of the two distinct liquid phases of water.
 
Our results suggest novel experiments. For instance the compressibility of
HDA should be very different depending on whether or not one anneals the
system after changing the pressure. The RP phase can be considered as the
real equation of state of the amorphous water (as suggested in
Ref.\cite{finney_vhda} for the VHDA at $p=11.5$ kbar).  Exploring the RP
curve is a challenge for experimentalists due to the instability towards
crystallization.  Just as at $p=0$ a continuum of structures could be
unveiled by cautious handling, it is quite possible that a similar feat can
be accomplished at $p \neq 0$ and the full RP line investigated, or at
least a large portion of it.

We should like to acknowledge stimulating discussions with M. Bernasconi,
V. Buch, M. Krack and M. L. Klein.


\begin{thebibliography}{99}
\bibitem{ft8} We used the Parrinello-Rahman MD method\cite{pr}, the
  Berendsen thermostat and a time step of 2 fs. Electrostatic interactions
  were treated by the particle-mesh Ewald method. The initial configuration
  was proton-disordered I$_{\rm h}$ ice with zero dipole moment.
\bibitem{ft1} All points in Fig.\ref{fig_ro_p} were equilibrated for 5 ns
  and averaged over 0.5 ns. Additional equilibration for 25 -- 100 ns was
  performed during annealing at $T=170$ K and also at $T=80$ K after
  decompression from 1.5 -- 6 kbar to $p = 0$.
\bibitem{ft4} Similar treatment was applied in Ref.\cite{loerting} at $p =
  8.4, 11$ and 19 kbar.
\bibitem{networks} One of several crystalline phases found upon annealing
  HDA' to $T=175$ K\cite{klotz_recryst} is ice XII which also has 7 and
  8-membered rings.  We do not find any indication of interpenetrating
  networks in our RP up to 22.5 kbar.
\end{thebibliography}
\end{document}